\documentclass[3p,twocolumn]{elsarticle}

\usepackage{amsmath}

\begin{document}

\biboptions{sort&compress}

\title{Free energy approach to micellization and aggregation: Equilibrium,
metastability, and kinetics}

\author[tauchem]{Haim Diamant}
\ead{hdiamant@tau.ac.il}
\ead[url]{http://www.tau.ac.il/$\sim$hdiamant}

\author[tauphys]{David Andelman}
\ead{andelman@post.tau.ac.il}
\ead[url]{http://www.tau.ac.il/$\sim$andelman}

\address[tauchem]{Raymond and Beverly Sackler School of Chemistry,
Tel Aviv University, Tel Aviv 6997801, Israel}

\address[tauphys]{Raymond and Beverly Sackler School of Physics
\& Astronomy, Tel Aviv University, Tel Aviv 6997801, Israel}

\begin{abstract}

We review a recently developed micellization theory, which is based on
a free-energy approach and offers several advantages over the
conventional one, based on mass action and rate equations. As all the
results are derived from a single free-energy expression, one can
adapt the theory to different scenarios by merely modifying the
initial expression. We present results concerning various features of
micellization out of equilibrium, such as the existence of metastable
aggregates (premicelles), micellar nucleation and growth, transient
aggregates, and final relaxation toward equilibrium. Several
predictions that await experimental investigation are discussed.

\end{abstract}

\maketitle

\section{Introduction}
\label{sec_intro}

Micellization\,---\,the self-assembly of amphiphilic molecules into
nano-scale aggregates in solution above a well-defined concentration
(the CMC)\,---\,is a well-studied phenomenon
\cite{TanfordBook,MittalBook,LindmanBook,IsraelachviliBook,BenShaulBook,RosenBook,ZanaBook}; arguably,
the science of micelles predates nano-science by several
decades. Theoretically, too, micellization is an old problem, dating
back to the recognition of the hydrophobic effect
\cite{Kauzmann1959,TanfordBook}. From the chemical-physics point of
view, it can be viewed as a restricted demixing process
\cite{Besseling1999}, where the growth of the new phase is terminated
at a finite characteristic size (``micro-phase separation'').

%As such,
%it is conceptually related to other finite-size phenomena, such as
%stabilization of nano-crystals and pattern formation.

The prevalent thermodynamic theory for micelles at equilibrium has
been the one by Israelachvili, Mitchell, and Ninham \cite{IMN}. It
combines mass-action thermodynamic considerations with geometrical
packing arguments, to account for the CMC and the aggregate shape and
size. Various extensions to this theory, incorporating additional
molecular details, were subsequently introduced (e.g.,
Ref.~\cite{Nagarajan1991}).

Concerning the kinetics of aggregation, the prevalent approach can be
traced back to Smoluchowski's classical theory of coagulation
\cite{Smoluchowski1916}, which is based on a set of reaction-rate
equations, the ``reactants'' being the various-sized aggregates. Its
application to surfactant micellization, progressing through monomeric
step-like growth/disintegration, is described by the Becker-D\"oring
equations \cite{Becker1935}\,---\,an infinite set of ordinary
differential kinetic equations, which is written for the
concentrations of aggregates, and whose linearization yields a discrete
spectrum of relaxation rates \cite{Babintsev2012}. This approach was
criticized for its restriction to single-monomer kinetics,
disregarding effects of micellar fusion and fission
\cite{Alsoufi2012,Griffiths2013,Pineiro2015}, and was argued to be
limited to cases of high CMC and small aggregation number
\cite{Griffiths2013}.

We begin the discussion with the basic ingredients common to any
theory of micellization. In its simplest form, a surfactant solution
is a binary mixture of surfactant and solvent. As such, it has three
intensive control parameters, e.g., the temperature $T$, pressure $p$,
and total volume fraction of surfactant $\Phi$, or, alternatively,
$T$, $p$, and the surfactant chemical potential $\mu$. The choice of
control parameters is immaterial if the solution is at thermal
equilibrium. Its kinetics, however, can strongly depend on the
specific constraints, e.g., whether the system is closed (fixed
$\Phi$) or open (fixed $\mu$). Given the three control parameters, the
system has as free variables the set of volume fractions of aggregates
containing $k$ solute molecules, $\{\Phi_k\}_{k=1,2,\ldots}$. During
the kinetics of micelle formation these variables are
time-dependent. (In more complex situations they may also be
space-dependent.) At equilibrium they attain the steady values
$\{\Phi_k\}=\{\Phi_k^{\rm eq}\}$. Various theories may differ in the
way these dynamic variables are derived. Once $\{\Phi_k\}$ are known
(either at or out-of-equilibrium), one can obtain the full
distribution of aggregate sizes.

The free-energy approach to micellization is centered on the
free-energy density of the solution, $F$. Considering a closed system,
$F=F(T,p,\Phi,\{\Phi_k\})$, subject to the constraint
$\sum_k\Phi_k=\Phi$. (From now on the dependence on $T$ and $p$ will
be omitted for brevity.) The equilibrium state, meta-stable states,
kinetic barriers, and time evolution of aggregation are obtained,
respectively, as the global minimum, local minima, maxima, and
time-dependent trajectories, along the multi-variable landscape
defined by $F$. Thus, apart from presenting an alternative description
of micellization, the free-energy formalism provides additional
information on such issues as the properties of metastable aggregates,
nucleation barriers, and relaxation processes. The aim of this
contribution is to present the essence and main findings of this
approach, as published in
Refs.~\cite{jpcb07,jpcb07err,jcp09,jpcb11}. Theories of similar spirit
were presented in Ref.~\cite{Maibaum2004} for the thermodynamics of
surfactant micelles, Ref.~\cite{Nyrkova2005} for the thermodynamics of
block copolymer micelles, and Ref.~\cite{Bhattacharjee2013} for
fluctuations, metastability, and kinetics close to the CMC.

\section{Free energy landscape}
\label{sec_freeenergy}

The free-energy landscape as defined above, $F(\Phi,\{\Phi_k\})$, is
in principle of infinite dimensions. The analysis is made tractable if
we assume that, for each thermodynamic state of the solution, the
distribution of aggregate size is either sharply unimodal, describing
only monomers, or sharply bimodal, describing monomers and micelles of
size $m$. Given the total surfactant volume fraction $\Phi$, this
leaves only two relevant variables, $\Phi_1$ and $m$. The volume
fractions of aggregates and solvent are given by
$\Phi_m=\Phi-\Phi_1$ and $1-\Phi$, respectively. 
%Note that, since $m$
%is treated as a variable, this simplification still allows to study
%details of the aggregation, such as the time evolution of the
%aggregate size or its statistical distribution at metastable states,
%via the dynamics or occupancy of the thermodynamic states
%$(\Phi_1,m)$.

Two simple ingredients are used for the formulation of $F$. The first is
the Flory-Huggins theory of solutions. The second is a single
phenomenological function, $u(m)$, which incorporates the detailed
properties of the specific surfactant molecule and accounts for the
free-energy gain of transferring that molecule from the
aqueous-solution environment into a micelle of size $m$. The only
requirement for $u(m)$ is that it should have a single maximum at a
certain value of $m$, to ensure the stability of finite aggregates
(i.e., to terminate the growth of the demixed phase). See
Refs.~\cite{Maibaum2004,jpcb07} for a specific choice of $u(m)$ and
its relation to the properties of the surfactant molecule. The
resulting free-energy density is
\begin{equation}
\label{F}
\begin{split}
  F(\Phi,\Phi_1,m) =& \frac{\Phi_1}{n}\ln\Phi_1
  + \frac{\Phi_m}{nm} [\ln\Phi_m - mu(m)] \\
  &+ (1-\Phi)\ln(1-\Phi).
\end{split}
\end{equation}
This function accounts for the entropy of mixing of the three species
(solvent, monomer, micelle), and the amphiphilic nature of the
surfactant (through the non-monotonous $u(m)$). We have simplified
this equation (the only equation in this article) and the formulae to
follow by expressing energies in terms of the thermal energy, $k_{\rm
  B}T$, and volumes in terms of the solvent molecular volume, $a^3$;
the surfactant molecular volume is taken to be $na^3$. All the results
presented below derive from Eq.~(\ref{F}) through simple mathematical
procedures whose details are found in
Refs.~\cite{jpcb07,jpcb07err,jcp09,jpcb11}.

\begin{figure}
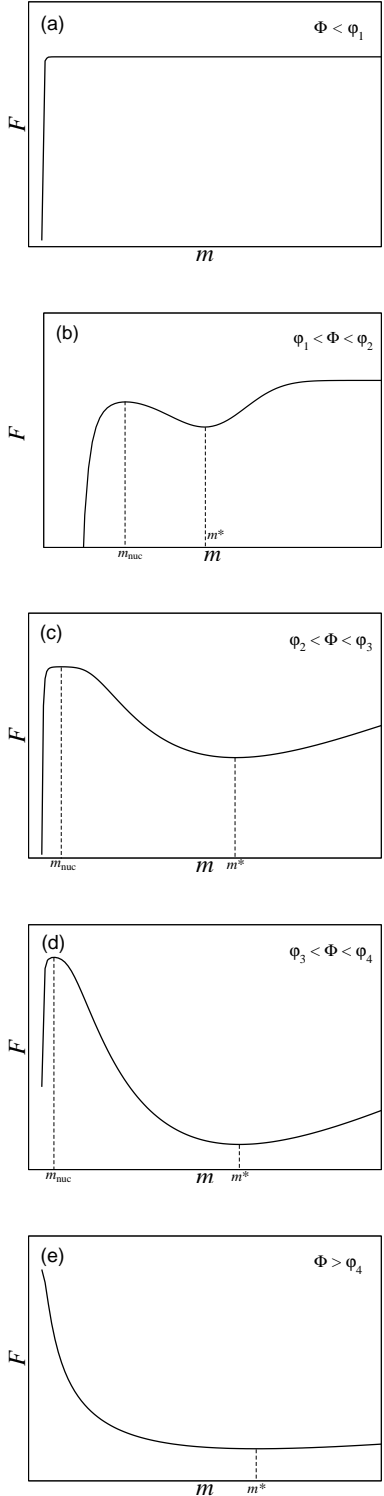

\centerline{\resizebox{0.3\textwidth}{!}
{\includegraphics{fig1a}}}
\vspace{0.63cm}
\centerline{\resizebox{0.3\textwidth}{!}
{\includegraphics{fig1b}}}
\vspace{0.63cm}
\centerline{\resizebox{0.3\textwidth}{!}
{\includegraphics{fig1c}}}
\vspace{0.63cm}
\centerline{\resizebox{0.3\textwidth}{!}
{\includegraphics{fig1d}}}
\vspace{0.63cm}
\centerline{\resizebox{0.3\textwidth}{!}
{\includegraphics{fig1e}}}
\caption{Schematic depiction of the free energy $F(\Phi_1^*(m),m)$ as
  a function of aggregation number $m$ in the different concentration
  regimes, as obtained from Eq.~(\ref{F}).}
\label{fig_F}
\end{figure}

As the total surfactant volume fraction $\Phi$ is increased, the shape
of the manifold defined by $F(\Phi_1,m)$ at fixed $\Phi$ changes,
revealing various regimes of aggregation, to be described below, and
as schematically shown in Fig.~\ref{fig_F}. The features seen in
Fig.~\ref{fig_F} (free-energy wells and barriers) are typically of
order $\sim 1$ $k_{\rm B}T$ per molecule, which amounts to tens of
$k_{\rm B}T$ per aggregate \cite{jpcb07,jpcb07err,jcp09,jpcb11}. In
the present review, which is focused on unifying underlying
mechanisms, we will not get into further numerical details concerning
specific systems. In the following discussion the term ``state''
refers to the {\em entire} solution, not to the state of the
surfactant molecules; thus, an ``aggregated state'' means a solution
containing both monomers and aggregates, with a given partitioning
between the two, $\Phi_1$ and $\Phi_m$, and with a given aggregate
size, $m$.

For any value of $\Phi$, $F$ of Eq.~(\ref{F}) has a single minimum
along the $\Phi_1$ axis, at $\Phi_1^*(\Phi,m)$. However, as long as
the surfactant volume fraction $\Phi$ is sufficiently low,
$F(\Phi_1^*,m)$ is monotonously increasing with $m$
(Fig.~\ref{fig_F}(a)), which implies a global minimum for a purely
monomeric solution (with $m=1$).

\section{Metastable premicelles and stable micelles}
\label{sec_metastable}

Above a certain volume fraction, $\Phi>\varphi_1$, $F(\Phi_1^*,m)$
becomes nonconvex as a function of $m$, and two extrema appear in
addition to the minimum at the monomeric state
(Fig.~\ref{fig_F}(b)). The first, which is a saddle point of
$F(\Phi_1,m)$, represents an unstable state containing critical nuclei
of size $m_{\rm nuc}$. The second, which is a local minimum of $F$, is
a metastable state with aggregates of size $m^*$.  In the analogy with
first-order phase transitions, $\varphi_1(T,p)$ corresponds to the
spinodal surface.

%% The values of $m_{\rm nuc}$ and $m^*$ are
%% given by the two solutions of the following equation \cite{jpcb11}:
%% \begin{equation}
%%   -\ln[\Phi-e^{-u(k)-mu\,'(k)-1+1/m}]/[m^2 u\,'(k)] = 1,
%% \label{mextrema}
%% \end{equation}
%% where $u'=du/dm$.  Once these aggregation numbers are found, the
%% volume fractions of monomers and aggregates are calculated according to
%% \begin{equation}
%%   \Phi_1^* = e^{-u(k)-mu\,'(k)-1+1/m}\ \ \
%%   \Phi_k^* = \Phi - \Phi_1^*.
%% \label{Phiextrema}
%% \end{equation}

Although the metastable state appears as soon as $\Phi>\varphi_1$,
closer inspection \cite{jpcb07} reveals that only above a higher
volume fraction, $\Phi>\varphi_2$ (Fig.~\ref{fig_F}(c)), does this
state become significantly occupied. In the range
$\varphi_1<\Phi<\varphi_2$ the value of $m^*$ increases, while the
fraction of surfactant molecules in the metastable aggregated state
remains negligible. For $\Phi>\varphi_2$, $m^*$ remains almost
constant, while the fraction of molecules in the metastable state
increases.

Above a higher total volume fraction, $\Phi>\varphi_3$, the aggregated
state $(\Phi_1^*,m^*)$ becomes the global free-energy minimum, and the
purely monomeric state turns metastable (Fig.~\ref{fig_F}(d)). In the
phase-transition analogy, $\varphi_3(T,p)$ is the binodal surface. The
value $\Phi=\varphi_3$ corresponds to the critical micelle
concentration (CMC) as it is commonly measured in experiments
\cite{jpcb07}. Thus, the range $\varphi_2 <\Phi<\varphi_3$ is
identified as the {\it premicellar regime}, and the metastable
aggregates are termed {\it premicelles}.

Finally, above a yet higher volume fraction, $\Phi>\varphi_4$, the
purely monomeric state becomes unstable, and the aggregated state
remains the sole free-energy minimum (Fig.~\ref{fig_F}(e)). In the
phase-transition analogy $\varphi_4(T,p)$ represents the second
spinodal surface of the mixture. We are not aware of an experiment in
which this latter change in solution behavior was observed.

Returning to the issue of metastable aggregates, their appearance may
be complicated by dynamic limitations. The results given above are
obtained under the assumption that the solution has indefinite time to
equilibrate. In practice there is a free-energy barrier,
$F(\Phi_1^*,m_{\rm nuc})$, to cross, which may take too long and
require the help of impurities to allow heterogeneous nucleation.
Another dynamic issue is the lifetime of the metastable premicelles
once they are formed. An analysis of the escape time from the
free-energy minimum $F(\Phi_1^*,m^*)$, across the barrier
$F(\Phi_1^*,m_{\rm nuc})$, back to the monomeric state, has been
performed based on Kramers' rate theory \cite{jcp09}. It showed that
reasonably long (say, longer than seconds) premicellar lifetimes are
obtained over a significant part of the premicellar
($\varphi_2<\Phi<\varphi_3$) region. In addition, the polydispersity
of premicelles was found to be only slightly larger than that of
stable micelles \cite{jcp09}.

Even after taking into account the limitations related to the
occupancy and lifetime of the metastable state, the theory predicts a
large extent of premicellar aggregation well below the CMC
\cite{jcp09}. Indeed, any sharp transition is smoothed by finite-size
effects, allowing the new state to be observed slightly below the
transition point \cite{HillBook}. However, the predicted extent of
premicellar aggregation can be an order of magnitude larger than what
would be expected from a simple finite-size correction. The difference
lies in the freedom to vary $m$ as compared to a simple two-state case
with monomers and aggregates of fixed $m$ \cite{jpcb07}.

\begin{figure}
\centerline{\resizebox{0.4\textwidth}{!}
{\includegraphics{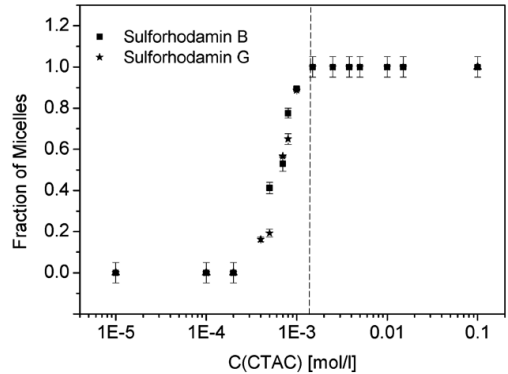}}}
\caption{Results of FCS measurements in cationic surfactant (CTAC)
  solution. The data points show the fraction of anionic dye molecules
  (sulforhodamin B or G), associated with aggregates, as a function of
  surfactant concentration. As surfactant aggregates form, they bind
  the oppositely charged dye molecules, consequently appearing in the
  FCS measurement as a an additional, slowly diffusing, fluorescent
  species. The vertical dashed line indicates the CMC as found from
  conductivity measurements in CTAC solutions containing sulforhodamin
  G, precluding a possible reduction of the actual CMC by the
  dye. Adapted with permission from Zettl et al.\ \cite{Zettl2005}. 
  %J.~Phys.~Chem.~B {\bf 109}, 13397--13401 (2005). 
  Copyright (2005) American Chemical Society.}
\label{fig_exp}
\end{figure}

Over the years, and especially in the past decade, there have been
quite a number of reports of premicelles, using a large variety of
experimental techniques\,---\,steady-state and time-resolved
fluorescence spectroscopy
\cite{Niu1992,Sahoo2002,Jaffer2008,Barnadas2009,Beija2010,Sakai2006,Choudhury2011,Sowmiya2010,Tiwari2011},
fluorescence correlation spectroscopy (FCS) \cite{Zettl2005}, UV
absorption spectroscopy \cite{Fu2015}, dielectric relaxation
spectroscopy \cite{Chen2011}, NMR \cite{Cui2008}, electrophoresis
\cite{Sabate2000}, and diffusion coefficient via radioactive labeling
\cite{Lindman1973}. As an example, Fig.~\ref{fig_exp} shows results of
FCS measurements in a cationic surfactant solution, revealing the
formation of aggregates at concentrations three times smaller than the
CMC \cite{Zettl2005}. Premicellar aggregation has been predicted also
by another thermodynamic model \cite{Bhattacharjee2013}, molecular
dynamics simulations \cite{LeBard2012}, as well as simulations of
idealized systems \cite{deMoraes2013}. Despite these numerous
indications, the controversy surrounding premicellar aggregation has
not been completely settled. (For alternative views, see
Refs.~\cite{Alsoufi2012,Pineiro2015}.) For example, since many
observations rely on fluorescence techniques, one should consider the
effect of the fluorescent dye on the micellization \cite{Freire2010}.

\section{Kinetics of aggregation}
\label{sec_kinetics}

The free-energy landscape, $F(\Phi_1,m)$, can be used also to study
kinetic pathways of the surfactant solution toward equilibrium. These
are derived as time-dependent trajectories on the surface
$(\Phi_1,m)$, determined by certain constraints. The single additional
input to the theory is a molecular time scale $\tau_0$. 
%(e.g., the
%typical time required for a surfactant molecule to diffuse along a
%distance comparable to its size).  
Thus, different aggregation processes can be treated on the same
footing. Another advantage of this approach is that, unlike models
based on the Becker-D\"oring scheme, the kinetics is not limited to
single-monomer steps.

The constraints that determine the trajectories become apparent if we
assume that the time scales of different aggregation stages are well
separated. Under this assumption, if we start from an
out-of-equilibrium monomeric solution, we generally find a three-stage
aggregation process, including slow nucleation, much faster growth,
and ultimate relaxation toward equilibrium. The constrained
trajectories depend also on the overall thermodynamic constraints
imposed on the solution, e.g., whether it is a closed system (fixed
$\Phi$) or an open one (fixed $\mu$), and may result in very
different kinetic pathways. Below we outline the constrained paths
which represent these stages, and the main results of this theory;
full details, along with numerical examples, are found in
Ref.~\cite{jpcb11}.

Consider a closed system with total surfactant volume fraction above
the CMC ($\Phi>\varphi_3$; Fig.~\ref{fig_F}(d)), starting from a
purely monomeric state and ending at the equilibrium state
$(\Phi_1^{\rm eq},m^{\rm eq})=(\Phi_1^*(m^*),m^*)$. The first stage is
an increase of the free energy from the metastable monomeric state to
the saddle point $F(\Phi_1^*(m_{\rm nuc}),m_{\rm nuc})$, i.e., the
formation of critical nuclei. Assuming that this slow activated
process satisfies quasi-equilibrium constrains the trajectory to
$(\Phi_1^*(m(t)),m(t))$, while $m(t)$ increases from $1$ to $m_{\rm
  nuc}$. The total nucleation time is $\tau_{\rm nuc}(\Phi)=\tau_0
e^{\Delta F_{\rm nuc}}$, where $\Delta F_{\rm nuc} \sim
F(\Phi_1^*(m_{\rm nuc}),m_{\rm nuc}) - F_1$ is the height of the
barrier per nucleus ($F_1$ being the free energy of the monomeric
state).

Various features of the nucleation stage can be calculated based on
this description \cite{jpcb11}. Taking the total surfactant volume
fraction $\Phi$ further above the CMC leads to a sharp decrease in
$\tau_{\rm nuc}$, a sharp increase in the concentration $c_{\rm nuc}$
of critical nuclei, and a gradual decrease in $m_{\rm nuc}$. 

In contrast, the results for the nucleation in an open system are
strikingly different. If transport of micelles from the bulk reservoir
is blocked or negligible, the solution is in contact with the
reservoir only through its monomeric concentration (the so-called
inter-micellar concentration), which hardly changes with further
increase of the bulk volume fraction above the CMC. Consequently, the
critical nuclei remain relatively rare and large, and the nucleation
barrier remains high. The resulting prediction is that homogeneous
nucleation of micelles in an open system should be kinetically
hindered.

The slow nucleation stage is followed by a much faster stage of
aggregate growth. Assuming that the growth is fed exclusively by
surrounding monomers implies that the number density of micelles,
$(\Phi-\Phi_1)/(na^3m)$, remains fixed and equal to $c_{\rm
  nuc}$. Thus, the appropriate constrained path representing this
stage is $(\Phi_1(t)=\Phi-na^3c_{\rm nuc}m(t),m(t))$. The rate of
growth may be limited either by the diffusion of monomers to the
aggregate or by the local kinetics at the aggregate. In the former
case, the growth is proportional to the spatial gradient of $\Phi_1$,
%$dm/dt\proptoD(\partial\Phi_1/\partial r)$, 
whereas in the latter it is proportional to the thermodynamic driving
force, i.e., the variation of $F$ with $m$ along the constrained path.
%$dm/dt\propto -\tau_0^{-1}(\delta F/\delta m)$.  
Analysis shows that both mechanisms may be relevant in practice
\cite{jpcb11}. The growth stage ends at the minimum of $F$ {\it along the
constrained path} defined above. In general this point on the landscape
does not coincide with the {\it global} minimum of $F$. Therefore, the
transient aggregate size reached at the end of the growth stage,
$\bar{m}$, may be either smaller or larger than the equilibrium size
$m^*$.

In the last stage of growth, the closed solution relaxes toward the
equilibrium state, $(\Phi_1^{\rm eq},m^{\rm
  eq})=(\Phi_1^*(m^*),m^*)$. Over this longer relaxation the
constraint on the number of micelles is lifted. At the same time,
nucleation or disintegration of entire micelles take too long. This
implies that the evolution of this final stage should occur through
micellar fusion (if $\bar{m}<m^*$) or fission (if $\bar{m}>m^*$). The
relaxation rate is related again to the thermodynamic driving force;
yet, in the latter stage it is given by the slope of $F$ along the
$[\phi_1^*(m(t)),m(t)]$ path toward equilibrium, without a constraint
on the concentration of micelles. In this final relaxation stage the
kinetics of an open system is again found to be strongly hindered.

Several kinetic characteristics described in this section have been
supported by experiments and other theories. Time-resolved small angle
x-ray scattering revealed the three stages presented above in
block-copolymer micellization \cite{Lund2013}. Three separate stages
were obtained also by another model based on kinetic equations
\cite{Neu2002}. The possibility of transient micelles relaxing into
micelles of different size was indicated by two other micellization
models \cite{Besseling1999,Griffiths2013}, as well as idealized
(two-dimensional) Monte-Carlo simulations \cite{deMoraes2013}. The
kinetic hindrance of micelle formation in open surfactant solutions is
supported by dialysis experiments, where the diffusive contact with
the reservoir does not allow the passage of micelles. The appearance
of micelles on the monomeric side was found to take hours
\cite{Morigaki2003}.

\section{Concluding remarks}
\label{sec_summary}

This short review has outlined the free-energy theoretical framework
that was recently developed for micellization. The formulation is
sufficiently general, in fact, to apply just as well to any
finite-size aggregation in solution, e.g., the formation of
surface-stabilized nanocrystals. This approach has several advantages,
such as its consistent and self-contained account of different
phenomena (all the results stemming from one free-energy function);
its simplicity, which allows to obtain many of the results
analytically and the rest by very basic numerics; and its easy
extension to other scenarios (by modifying that single function). On
the other hand, it should be kept in mind that the theory provides
only a crude deterministic description of much more complicated
stochastic phenomena. In particular, the assumption of a sharply
bimodal distribution of aggregate sizes should be relaxed to obtain a
reliable quantitative account of stages that involve crossing
a nucleation barrier.

We would like to highlight a few experimental implications. The
existence of metastable premicelles well below the CMC, as implied by
the theory, has received significant experimental support (see
Sec.~\ref{sec_metastable}), but is not considered as settled. Our
present point of view is that intrinsic homogeneous nucleation of
premicelles is kinetically suppressed and probably negligible. Their
observation requires, therefore, heterogeneous nucleation facilitated
by impurities \cite{jpcb07err}. Indeed, this sensitivity of
premicellar nucleation is a possible explanation for conflicting
experimental results. The potential appearance of transient micelles
larger than their equilibrium size is another strong prediction,
shared by other theories \cite{Besseling1999,Griffiths2013}, which to
our knowledge has not been checked experimentally. Another prediction,
which calls for more controlled experiments, is the very long kinetic
suppression of micelle nucleation in open systems whose exchange with
the reservoir is limited to monomers.

Finally, an interesting and unexplored aspect of micellization is the
dynamics following a deep quench beyond the ``spinodal'' $\varphi_4$
(see Fig.~\ref{fig_F}(e)). Comparison to the well-studied spinodal
decomposition of ordinary mixtures might underline the similarity and
difference between micellization and demixing phase transitions.

\vspace{.5cm}
\noindent
{\bf Acknowledgments}\\ We thank Martin Lenz for a helpful
discussion. DA acknowledges support from the ISF-NSFC joint research
program (Grant No. 885/15), the Israel Science Foundation (ISF) (Grant
No. 438/12), and the United States--Israel Binational Science
Foundation (BSF) (Grant No. 2012/060).

\end{document}